# SYSTEMATIC EXPANSION FOR FULL QCD BASED ON THE VALENCE APPROXIMATION

J. Sexton[1], and D. Weingarten
IBM Research
P.O. Box 218, Yorktown Heights, NY 10598

## ABSTRACT

We construct a systematic expansion for full QCD. The leading term gives the valence (quenched) approximation.

---

[1]permanent address: Department of Mathematics, Trinity College, Dublin 2, Republic of Ireland

# 1 INTRODUCTION

Several predictions obtained recently in the valence (quenched) approximation to the infinite volume, continuum limit of lattice QCD lie not far from experiment. For low lying hadron masses [1], valence approximation results are within 6% ± 8% of experiment. For decay constants [2] the valence approximation differs from experiment by increments ranging from 12% ± 11% to 17% ± 6%. Missing from these calculations, however, is an independent theoretical estimate of the error arising from the valence approximation.

In the present article, we develop a systematic expansion for lattice QCD including the full effect of quark vacuum polarization. The leading term in this scheme is the valence approximation. If an infinite collection of higher terms is taken into account, full QCD is reproduced exactly. We then derive a formula which can be used to estimate the error in any vacuum expectation value obtained by truncation of this expansion to some finite number of terms. Our expansion assumes quarks occur in pairs with equal mass.

In an exact treatment of QCD, virtual quark-antiquark pairs produced by a chromoelectric field reduce the field's intensity by a factor which depends both on the field's momentum and on its intensity. In the valence approximation this factor, analogous to a dielectric constant, is approximated by its zero-field-momentum zero-field-intensity limit [3]. The approximation which we consider here may be pictured as incorporating an inverse dielectric constant which is a sum of terms which progressively more accurately reproduce the correct dependence of the inverse dielectric constant on field momentum and field intensity.

In Section 2 we introduce definitions. In Section 3, we construct an expansion for the dependence of vacuum polarization on field momentum and field strength and derive an expression for the error in any vacuum expectation value arising from a truncation of the expansion. In Sections 4 and 5 two variations on the expansion of Section 3 are considered. In Sections 6, 7, and 8 we present an algorithm for evaluating the terms in these expansion. In Section 9, we describe a trial calculation using the expansion in Section 3. In Appendices A, B, C and D, we prove convergence of the sequence of vacuum expectation values arising from truncated forms of our expansions. Appendix E gives a calculation of a set of parameters needed by the algorithm in Section 6.

The main motivation for the present work is to find a way of determining directly from QCD the errors arising in valence approximation calculations of hadron masses and decay constants. This goal is accomplished in principle by the error formula for our expansion truncated after the leading term. A crucial question which we have not yet answered, however, is whether in practice the determination of valence approximation errors using our algorithm would be any faster than a direct comparison of valence approximation results with numbers found by the best present algorithms for full QCD. Whether or not the method we propose turns out to be useful in practice for quantitative error estimates, it appears to us that it does help provide a qualitative understanding of the mechanism underlying the relatively close agreement found in Refs. [1, 2] between valence approximation predictions and the real world. In particular, the results we present in Section 9 are clear evidence of at least one set of parameters and expectation values for which the main effect of vacuum polarization is absorbed by the dielectric constant implicit in the valence approximation.

We are aware of two other strategies for evaluating the relation between the valence approximation and full QCD. The application of chiral perturbation theory to estimating the errors introduced by the valence approximation has been considered by several groups [4, 5, 6]. The limiting behavior at small quark mass of a variety of predictions of the valence approximation has been shown to be qualitatively different from the behavior of full QCD. The quark mass below which these difficulties become quantitatively significant, however, appears to be well below the average of the up and down quark masses [7, 8]. For physical values of quark mass, several unknown parameters enter chiral perturbation theory predictions of the errors in most valence approximation results. Quantitative determination of these errors is therefore not possible at present. Another method for evaluating of the effect of virtual quark-antiquark pair production on QCD predictions is discussed in Ref. [9]. This calculation uses a weak coupling expansion to leading order and is valid for small values of the gauge coupling constant and large values of the quark mass. The results we report in our trial calculation in Section 9 are qualitatively consistent with those described in Ref. [9]. Reviews of a variety of other recent valence approximation calculations are given in Ref. [10].

## 2  DEFINITIONS

We consider Wilson's formulation of Euclidean QCD on some finite lattice. A lattice gauge field consists of an assignment of an element $u(x_1, x_2)$ of the fundamental representation of $SU(3)$ to each oriented nearest neighbor pair of sites $(x_1, x_2)$ with the usual restriction that $u(x_1, x_2)$ is the adjoint $u(x_2, x_1)^\dagger$.

Define the Hilbert space $\mathcal{F}$ to consist of complex valued functions $f$ of the lattice gauge fields with finite value of the norm

$$\|f\|^2 = \zeta^{-1} \int d\mu \, |f|^2 \, exp(\mathcal{S}), \qquad (2.1)$$
$$\zeta = \int d\mu \, exp(\mathcal{S}).$$

The inner product on $\mathcal{F}$ is

$$(f, f') = \zeta^{-1} \int d\mu \, f^* \, f' \, exp(\mathcal{S}), \qquad (2.2)$$
$$\zeta = \int d\mu \, exp(\mathcal{S}).$$

Here $\mathcal{S}$ is some real-valued function of the field which is bounded in absolute value. Several different possible choices of $\mathcal{S}$ will be discussed in Sections 3 and 4. A linearly independent basis for $\mathcal{F}$ consists of the collection $\{f_i\}$ of all possible products of matrix elements of irreduceable representations of $SU(3)$ including exactly one matrix element for each link, with links differing only by a flip of orientation identified. Distinct $f_i$ are then orthogonal with respect to the inner product of Eq. (2.2) for $\mathcal{S}$ of 0. For convenience in Appendix C, we choose the $f_i$ to be normalized with respect to the inner product with $\mathcal{S}$ of 0.

Let $d_i$ be the sum over all links of the dimension of the $SU(3)$ representation assigned to that link by $f_i$. We assume the sequence $\{f_i\}$ is ordered in such a way that $d_i$ is a nondecreasing function of $i$. Applying a Gram-Schmidt process to $\{f_i\}$ using the inner product of Eq. (2.2) for some nonzero choice of $\mathcal{S}$ gives an orthonormal basis $\{\hat{f}_i\}$ for $\mathcal{F}$.

Although the expansion to be constructed in Section 3 can be defined using only $\mathcal{F}$, for purposes of constructing an algorithm to evaluate this expansion it is slightly more convenient to work with the subspace $\mathcal{H}$ of $\mathcal{F}$ which is invariant under all lattice translations, rotations, reflections, gauge transforms and complex conjugation. Let $h_i$ be the projection of $f_i$ onto $\mathcal{H}$. Since rotation, translation, reflection, gauge transformation and complex conjugation leave the value of $d_i$ unchanged, $h_i$ will be a linear combination of a collection of $f_j$ all of which have $d_j$ equal to $d_i$. Most $h_i$ obtained in this way will be linearly dependent on the set of $h_j$ with $j < i$. Working upwards from $i$ of 0, we eliminate any $h_i$ which is dependent on surviving $h_j$ with $j < i$. A Gram-Schmidt process on the surviving sequence gives an orthonormal basis $\{\hat{h}_i\}$ for $\mathcal{H}$.

Typical vectors in $\mathcal{H}$ are the function with value 1 for all gauge fields and the Wilson plaquette action

$$P \;=\; \sum_{(x_1,\ldots x_4)} tr[u(x_1,x_2)u(x_2,x_3)u(x_3,x_4)u(x_4,x_1)], \qquad (2.3)$$

where the sum is taken over all oriented plaquettes $(x_1,\ldots x_4)$ consisting of sequences of four successive nearest neighbors, with sequences related by a cyclic permutation identified. Any sum of traces of products of $u(x,y)$ over all rotations, translations, reflections and order reversals of some closed path gives yet another element of $\mathcal{H}$. The basis vector $\hat{h}_0$ is the function with constant value 1, and $\hat{h}_1$ is the normalized projection of $P$ orthogonal to $\hat{h}_0$. The basis vectors $\hat{h}_2$, $\hat{h}_3$, and $\hat{h}_4$, are each found by continuing the Gram-Schmidt process on the three different sums of traces of products of $u(x,y)$ along one of the three distinct shapes of closed paths consisting of six lattice links.

We now define the lattice vacuum expectation value. We assume quarks occur in degenerate pairs for some set of masses strictly greater than 0. Let $M$ be Wilson's coupling matrix among half the quark fields, one from each degenerate pair. We impose periodic boundary conditions. For any function of the gauge fields $G$ with bounded absolute value, a regulated form of the vacuum expectation value obtained after integrating out quark fields is

$$< G >_R = Z^{-1} \int d\mu \, G \, det(M^\dagger M + R) \, exp[\frac{\beta}{6} P],$$
$$Z_R = \int d\mu \, det(M^\dagger M + R) \, exp(\frac{\beta}{6} P), \qquad (2.4)$$

where $\beta$ is $6/g^2$ for bare gauge coupling constant $g$, $\mu$ is the product of one copy of $SU(3)$ Haar measure for each link variable on the lattice, and $R$ a small nonnegative parameter. The extension of Eq. (2.4) to vacuum expectations of products of quark and antiquark fields is not needed for the present discussion and will be omitted for simplicity. Since the Wilson coupling matrix $M$ obeys

$$det(M) = det(M^\dagger) \qquad (2.5)$$

the expectation $< G >_0$ is the usual vacuum expectation of lattice QCD. For any bounded $G$, $< G >_R$ approaches $< G >_0$ as $R$ goes to 0.

We introduce the regulator $R$ in the definition of $< G >_R$ to provide a mathematically convenient rule for handling rare gauge configuration on which $M$ becomes singular in the valence approximation. Monte Carlo valence approximation calculations often find averages of quantities involving $M^{-1}$ at values of the quark mass for which some configurations exist, such as all link variables close to the identity matrix, for which $M$ has eigenvalues arbitrarily close to 0. These configurations are not encountered in practice because their total weight within the path integral is extremely small. It is generally believed that for any positive choice of quark masses, the total valence approximation measure of configurations with minimal $M^\dagger M$ eigenvalue below $O(m_q^2)$ goes to zero very rapidly in the limit of large lattice volume with lattice spacing held fixed. The expansion to be considered below will be done with some nonzero value of $R$ much smaller than $O(m_q^2)$. After taking a limit of infinite volume of any vacuum expectation, a limit of zero $R$ should leave the result essentially unchanged. For notational simplicity, the $R$ subscript will be deleted from $< G >_R$ and $Z_R$ in the following.

The regulation parameter $R$ is needed only for the Wilson quark coupling martix. For Kogut-Susskind quarks the spectrum of $M^\dagger M$ is bounded from below by $m_q^2$ for all gauge configurations.

## 3 EXPANSION

For any choice of $\mathcal{S}$, bounded in absolute value, in Eq. (2.2), the space $\mathcal{H}$ can be used to construct an expansion for $\log det(M^\dagger M + R)$. Since $M$ is a finite matrix with matrix elements bounded uniformly over all gauge fields, the spectrum of $M^\dagger M + R$ is bounded from above by some constant $A$. Thus $det(M^\dagger M + R)$ has a finite value of the norm defined by Eq. (2.1) and is in $\mathcal{F}$. In addition $det(M^\dagger M + R)$ is real-valued and rotation, translation, reflection and gauge invariant. It is therefore in $\mathcal{H}$.

Since the spectrum of $M^\dagger M + R$ is bounded from below by $R$, $det(M^\dagger M + R)$ is bounded from both above and below. Thus $\| \log det(M^\dagger M + R) \|$ is finite and $\log det(M^\dagger M + R)$ is also in $\mathcal{H}$. Using the orthonormal basis $\{\hat{h}_i\}$ of $\mathcal{H}$, we therefore have the convergent expansion

$$\log det(M^\dagger M + R) = \sum_i a_i \hat{h}_i, \qquad (3.1)$$

$$a_i = (\hat{h}_i, \log det(M^\dagger M + R)). \qquad (3.2)$$

An algorithm for the numerical evaluation of the coefficients $a_i$ is presented in Section 6.

For any $G$ bounded in absolute value, an approximation to $< G >$ can be obtained by combining Eqs. (3.1) and (3.2) with Eq. (2.4). The expectation value defined by Eq. (2.4) can be reexpressed

$$< G > = Z^{-1} \int d\mu\, G\, exp[L_n + Q_n + \frac{\beta}{6} P],$$

$$Z = \int d\mu\, exp[L_n + Q_n + \frac{\beta}{6} P], \qquad (3.3)$$

where $n$ is some positive integer and the partial sum $L_n$ and remainder $Q_n$ are

$$L_n = \sum_{i=0}^{n} a_i \hat{h}_i, \qquad (3.4)$$

$$Q_n = \log det(M^\dagger M + R) - L_n,$$

$$= \sum_{i>n} a_i \hat{h}_i. \qquad (3.5)$$

As $n$ becomes large, Eq. (3.1) implies $Q_n$ approaches 0 in $\mathcal{F}$. Thus it appears reasonable to try to approximate Eqs. (3.3) by omitting $Q_n$. We obtain

$$<G>_n = Z_n^{-1} \int d\mu\, G\, exp[L_n + \frac{\beta}{6}P],$$

$$Z_n = \int d\mu\, exp[L_n + \frac{\beta}{6}P]. \qquad (3.6)$$

The expectation $<G>_0$ is pure QCD with the quark determinant simply removed and no shift in $\beta$. The expectation $<G>_1$ is the valence approximation including a shift in $\beta$. For any $G$ bounded in absolute value, the approximate expectation $<G>_n$ approaches $<G>$ as $n$ becomes large. A proof is given in Appendices A, B, C and D.

The asymptotic expansion to leading order in $Q_n$ for the error in $<G>_n$ is the correlation

$$<G> - <G>_n = <(Q_n - <Q_n>_n)(G - <G>_n)>_n . \qquad (3.7)$$

Combining Eq. (3.7) and the Cauchy-Schwarz inequality we can obtain the bound

$$(<G> - <G>_n)^2 \leq <(Q_n - <Q_n>_n)^2>_n <(G - <G>_n)^2>_n . \qquad (3.8)$$

If the estimate for $<G> - <G>_n$ in Eq. (3.7) turns out to be small in comparision to $<G>_n$, the approximation $<G>_n$ and the error estimate $<G>$ $-<G>_n$ should both be reliable. Conversely if the error given by Eq. (3.7) is significant in comparision to $<G>_n$, neither the approximate expectation $<G>_n$ nor $<G> - <G>_n$ can be considered reliable. Seen through the eyes of weak coupling perturbation theory, Eq. (3.7) gives the one quark loop approximation to the error $<G> - <G>_n$. Thus the reliability of this error estimate is not directly related to the rate of convergence of the sequence of $<G>_n$ as $n$ is made large. Higher order terms in the loop expansion for $<G> - <G>_n$ will be discussed in slightly more detail in Section 5.1.

To try to minimize the valence approximation error $<G> - <G>_1$, we now take $\mathcal{S}$ in Eq. (2.2) to be the effective action entering Eq. (3.6) for $<G>_1$,

$$\begin{aligned} \mathcal{S} &= L_1 + \frac{\beta}{6} P, \\ &= \alpha + \frac{\beta'}{6} P, \end{aligned} \quad (3.9)$$

with $\alpha$ and $\beta'$ given by

$$\begin{aligned} \alpha &= a_0 + \frac{a_1 <P>_1}{\sqrt{<[P- <P>_1]^2>}}, \\ \beta' &= \beta + \Delta\beta, \\ \Delta\beta &= \frac{6 a_1}{\sqrt{<[P- <P>_1]^2>}}. \end{aligned} \quad (3.10)$$

If $\mathcal{S}$ is held fixed in the inner product $(\ldots,\ldots)$ and in the definition of the vacuum expectation $<\ldots>_1$ but $a_0$ and $a_1$ are varied in $Q_1$, the values of $a_0$ and $a_1$ given by Eq. (3.2) using $(\ldots,\ldots)$ with $\mathcal{S}$ of Eq. (3.9) minimize $<(Q_1 - <Q_1>_1)^2>_1$. Thus by Eq. (3.8) the error $(<G> - <G>_1)^2$ should tend to be minimized by the choice Eq. (3.9).

For any choice of $\beta$, the valence approximation value $\beta'$ is to be found by solving

$$\beta' = \beta + \Delta\beta(\beta'). \quad (3.11)$$

With any reasonable number of flavors of quarks, less than 10 for example, it is easily confirmed numerically that $\Delta\beta(\beta')$ is a monotone increasing function of $\beta'$ with derivative significantly less than 1. For two flavors of quarks with $k$ of 0.1600 considered in Section 9, we found that $\Delta\beta$ rises from below 0.1 at $\beta'$ of 0 to below 0.3 at $\beta'$ near 6.0. Eq. (3.11) can then be solved by a fixed point iteration taking $\beta$ as an initial value of $\beta'$. Alternatively, if the valence approximation $\beta'$ is chosen first, Eq. (3.11) gives directly the corresponding $\beta$ for full QCD. This procedure will be adopted in the example to be discussed in Section 9.

It may be useful to mention here that the hopping constant expansion for $\log det(M^\dagger M)$ expresses this quantity as a linear combination of Wilson loops formally similar to Eq. (3.1) and can be used to obtain an approximation to $<G>$ similar to Eq. (3.6). In two crucial ways, however, the expansion of Eq. (3.1) differs from the hopping constant expansion, and approximation Eq. (3.6) differs from the corresponding approximation using the hopping constant expansion. First, the validity of expansion Eq. (3.1) and the accuracy of approximation Eq. (3.6) are not restricted to the range of large quark mass to which the hopping constant expansion and its related approximation apply. As we have already shown and as will be discussed in Appendices A, B, C and D, expansion Eq. (3.1) and approximation Eq. (3.6) apply as long as the quark mass is greater than 0. Second, even for values of quark mass at which the hopping constant expansion does converge, Eqs. (3.1) and (3.2) differ from the hopping constant expansion by an infinite rearrangement. That is each term which appears in Eqs. (3.1) and (3.2) is a linear combination of an infinite set of the terms appearing in the hopping constant expansion, and vice versa.

A wide range of other possible expansions and approximate vacuum expectation values similar to Eqs. (3.1) and (3.6), respectively, can be constructed by changing the choice of $\mathcal{S}$ and expansion basis $\{\hat{h}_i\}$. Some of these will be discussed in Section 4. Yet another class of possibilities, which we will not discuss here in detail, is to choose the $a_i$ in Eq. (3.1) to force the first order shift in Eq. (3.7) to zero for a particular $G$ or set of $G$, such as the pion propagator or the rho propagator.

## 4 OPTIMIZED EXPANSION

In Section 3, we showed that the leading contribution to a valence approximation error $(<G> - <G>_1)^2$ will tend to be minimized by choosing the effective action $\mathcal{S}$ in Eq. (2.2) to be the effective action $L_1 + \frac{\beta}{6}P$ occurring in the valence approximation vacuum expectation $<\ldots>_1$. Since the effective action contribution $L_1$ depends on the choice of $\mathcal{S}$, we are then led to a nonlinear equation Eq. (3.11) to be solved to find $\mathcal{S}$. A similar recursive definition of $\mathcal{S}$ as $L_n + \frac{\beta}{6}P$ will tend to minimize the leading contribution to the error $(<G> - <G>_n)^2$. This choice gives a more complicated version of Eq. (3.11) to be solved for the coefficients entering $\mathcal{S}$.

In Section 3 we also argued, however, that the function $\Delta\beta(\beta)$ defined by Eq. (3.10) has a small derivative. This observation can roughly be rephrased as saying that in the neighborhood of our choice of $\mathcal{S}$, the valence approximation effective action coefficients vary only slowly with respect to changes in the $\mathcal{S}$ entering Eq. (2.2) defining $(\ldots,\ldots)$. We would expect similarly weak dependence on $\mathcal{S}$ of the effective action $L_n + \frac{\beta}{6}P$ of higher order vacuum expectations $< \ldots >_n$. Thus the improvement obtained in any $(<G> - <G>_n)^2$ by using a complicated $\mathcal{S}$ in Eq. (2.2) in place of the $\mathcal{S}$ of Eq. (3.9) may not be very large. Similarly, an $\mathcal{S}$ which is a close approximation to $L_n + \frac{\beta}{6}P$ may provide very nearly all of the decrease available in a typical $(<G> - <G>_n)^2$ by optimal choice of $\mathcal{S}$. Rather than trying to solve a full recursive equation to optimize $\mathcal{S}$ for $< \ldots >_n$, a choice close to optimal is probably

$$\mathcal{S}_n = L_n + \frac{\beta}{6}P, \tag{4.1}$$

with $L_n$ itself taken from Eq. (3.4) using $\mathcal{S}$ of Eq. (3.9). A choice yet closer to optimal would be to apply Eq. (4.1) iteratively taking $L_n$ from an inner product defined using $\mathcal{S}_{n-1}$.

## 5 YET ANOTHER EXPANSION

If $<G>$ in Eq. (3.3) is expanded as a power series in $Q_n$, $<G>_n$ of Eq. (3.6) gives the constant term, and $<G> - <G>_n$ of Eq. (3.7) gives the linear term. Continuing to expand $<G>$ in powers of $Q_n$ yields an infinite series. It is likely that the resulting series can not be summed and is only an asymptotic expansion in small $Q_n$. An alternate expansion for $<G>$ as a ratio of power series in $Q_n$, however, can be proved convergent. The leading approximation to this ratio also reduces to $<G>_n$ plus the first correction in Eq. (3.7).

Expanding the two exponentials which appear in Eq. (3.3) as power series in $Q_n$ gives

$$\begin{aligned} <G> &= \frac{Y}{Z}, \\ Y &= \sum_m \frac{<G(Q_n)^m>_n}{m!}, \\ Z &= \sum_m \frac{<(Q_n)^m>_n}{m!}. \end{aligned} \quad (5.1)$$

The series in Eq. (5.1) are both convergent since $Q_n$, for any choice of $n$, is bounded in absolute value by a constant independent of the gauge field. This bound on $Q_n$ holds since both $\log det(M^\dagger M + R)$ and $L_n$ in Eq. (3.5) defining $Q_n$ obey such bounds. The algorithm to be described in Section 6 can be adapted to evaluate each of the individual terms in the series in Eq. (5.1). Evaluating terms beyond $m$ of 1 using this method, however, is probably very time consuming.

The terms of order $m$ in Eq. (5.1) give, roughly, the $m$ quark loop contribution to the error in $<G>_n$. The truncation of the sums in Eq. (5.1) yields a two parameter family of possible approximations to $<G>$. For $n$ and $m$ both 1, $<Q_1>_1$ vanishes so that Eq. (5.1) is equivalent to adding to $<G>_1$ the error estimate of Eq. (3.7).

## 6 TRACE LOG ALGORITHM

The quantity $\log det(M^\dagger M + R)$ needed for Eq. (3.2) obeys the identity

$$\log det(M^\dagger M + R) = tr \log(M^\dagger M + R). \quad (6.1)$$

We now consider an algorithm for finding $tr \log(M^\dagger M + R)$. The algorithm exploits properties of the Chebyshev polynomials. Combined with Eq. (3.2), this algorithm gives the coefficients $a_i$. As discussed in Section 2, we will assume the quark masses, lattice volume, and Monte Carlo ensemble size have been chosen in such a way that for all gauge configurations encountered the minimal eigenvalue $B$ of $M^\dagger M$ lies well above $R$. The effect of $R$ in $tr \log(M^\dagger M + R)$ is then negligable, and we omit $R$ in the remainder of the paper.

To evaluate $tr \log(M^\dagger M)$ we begin by generating an ensemble of gaussian random complex-valued pseudo-quark fields $\phi_i(x)$, where $i$ is a multi-index ranging over all combinations of quark spin, color and flavor and $x$ ranges over lattice sites. For each $i$ and $x$ we choose $\phi_i(x)$ to be an independent random variable such that the average over this ensemble $<<\ldots>>$ gives

$$<<\phi_i(x)\phi_j(y)>>= 0,$$
$$<<\phi_i^*(x)\phi_j(y)>>= \delta_{ij}\delta_{xy}. \qquad (6.2)$$

We then have

$$tr\log(M^\dagger M) \;=\; <<((\phi, \log(M^\dagger M)\phi))>>, \qquad (6.3)$$

where $((\ldots,\ldots))$ is the inner product on the vector space of pseudo-quark fields

$$((f,g)) = \sum_{ix} f_i^*(x)\,g_i(x). \qquad (6.4)$$

Finding the inner product of two such vectors requires a comparatively small amount of arithmetic. The problem of evaluating the trace $tr\log(M^\dagger M)$ is thus reduced to finding $\log(M^\dagger M)\phi$ for a large ensemble of $\phi$.

For the evaluation of $\log(M^\dagger M)\phi$ we combine properties of the Chebyshev polynomials with the restriction that the eigenvalues of $M^\dagger M$ lie between upper and lower bounds $A$ and $B$, respectively. Define the operator $Y$ and the parameter $\epsilon$ to be

$$Y \;=\; \frac{M^\dagger M}{A}, \qquad (6.5)$$
$$\epsilon \;=\; \frac{B}{A}. \qquad (6.6)$$

In Appendix E, we will show that for any $n$ greater than 1 there are a set of coefficients $b_i$, such that for any number $y$ between $\epsilon$ and 1

$$\log y \;=\; \sum_{i=0}^{n} b_i T_i^*\left(\frac{1-y}{1-\epsilon}\right) + \delta \log y, \qquad (6.7)$$
$$|\delta| \;<\; 2\,exp(-2n\sqrt{\epsilon}), \qquad (6.8)$$

where the $T_i^*$ are Chebyshev polynomials. For large values of $n$, the inequality of Eq. (6.8) is nearly saturated. Since $Y$ is a self-adjoint operator with all eigenvalues between $\epsilon$ and 1, for any vector $\phi$

$$\log(Y)\phi = \sum_{i=0}^{n} b_i T_i^*(\frac{1-Y}{1-\epsilon})\phi + \delta \log(Y)\phi, \tag{6.9}$$

with $\delta$ bounded according to Eq. (6.8).

An iterative algorithm to evaluate the sum in Eq. (6.9) can be obtained from the recursion relation

$$T_{i+1}^*(z) = (4z - 2)T_i^*(z) - T_{i-1}^*(z) \tag{6.10}$$

and initial values

$$T_0^*(z) = 1,$$
$$T_{-1}^*(z) = 2z - 1. \tag{6.11}$$

Define the sequences $C_i$ and $D_i$ for $0 < i \leq n$ by

$$C_i = T_0^*(Y)\phi + c_i(\frac{2 + 2\epsilon - 4Y}{1 - \epsilon}C_{i-1} - D_{i-1})\phi,$$
$$D_i = T_{-1}^*(Y)\phi + c_i C_{i-1}\phi, \tag{6.12}$$

with initial values

$$C_0 = T_0^*(Y)\phi,$$
$$D_0 = T_{-1}^*(Y)\phi. \tag{6.13}$$

The coefficients $c_i$ in Eq. (6.12) are found from the $b_i$ in Eqs. (6.7) and (6.9) by

$$c_i = \frac{b_{n+1-i}}{b_{n-i}}. \tag{6.14}$$

Eqs. (6.10 - 6.14) imply that $b_0 C_n$ gives the sum in Eq. (6.9) and is therefore an approximation to $\log(Y)\phi$ with relative error less than $|\delta|$.

The final result, by Eqs. (6.3) and (6.5), is

$$tr \log(M^\dagger M) = \;<< ((\phi, b_0 C_n)) >> + d_M \log(A), \tag{6.15}$$

where $d_M$ is the dimension of the matrix $M$.

# 7 MAXIMUM AND MINIMUM EIGENVALUES

Since for small values of $\delta$, the number of iterations required to obtain a fixed value of $\delta$ in Eq. (6.7) becomes a linear function of $\sqrt{A/B}$, the optimal choices for $A$ and $B$ become $\lambda_{max}$ and $\lambda_{min}$, respectively.

An efficient algorithm [12] for estimating the maximum and minimum eigenvalues of $M^\dagger M$ uses the Lanczos method to construct a tridiagonal approximation to $M^\dagger M$. Define the sequences of real numbers $\alpha_1, \ldots \alpha_m$ and $\beta_0, \ldots \beta_m$ and the sequences of pseudo-quark fields $q_0, \ldots q_m$ and $r_0, \ldots r_m$ by

$$\begin{align}
q_{i+1} &= (\beta_i)^{-1} r_i, \\
\alpha_{i+1} &= ((q_{i+1}, M^\dagger M q_{i+1})), \\
r^{i+1} &= M^\dagger M q^{i+1} - \alpha_{i+1} q_{i+1} - \beta_i q_i, \\
\beta_{i+1} &= \sqrt{((r_{i+1}, r_{i+1}))},
\end{align} \tag{7.1}$$

with $\beta_0$ of 1, $q_0$ identically 0, and $r_0$ a randomly chosen pseudo-quark field with norm 1. Here $m$ is the number of distinct eigenvalues of $M^\dagger M$.

It can be shown that the sequence of pseudo-quark fields $q_1, \ldots q_m$ generated by Eq. (7.1) is orthonormal, and the space spanned by these vectors is invariant under the action of $M^\dagger M$. The space spanned by $q_1, \ldots q_m$ is smaller than the whole space of pseudo-quark fields only if one or more of the eigenvalues of $M^\dagger M$ is degenerate. In the basis $q_1, \ldots q_m$, $M^\dagger M$ is tridiagonal. All matrix elements of $T_{ij}$

$$T_{ij} = ((q_i, M^\dagger M q_j)) \tag{7.2}$$

vanish except

$$\begin{align}
T_{i-1\,i} &= \beta_{i-1}, \\
T_{ii} &= \alpha_i, \\
T_{i+1\,i} &= \beta_i.
\end{align} \tag{7.3}$$

Each distinct eigenvalue of $M^\dagger M$ occurs exactly once as an eigenvalue of the matrix of Eq. (7.3). As $n$ grows, the maximum and minimum eigenvalues $\lambda_{max}^n$ and $\lambda^n_{min}$ of the submatrix $T^n_{ij}$ with $1 \leq i,j \leq n$ approach the true maximum and minimum eigenvalues $\lambda_{max}$ and $\lambda_{min}$ of $M^\dagger M$ with errors falling exponentially in $n$.

To extract $\lambda^n_{max}$ and $\lambda^n_{min}$ from $T^n$, define the polynomial $p^n(\lambda)$ to be the determinant

$$p^n(\lambda) = det(T^n - \lambda I^n), \tag{7.4}$$

where $I^n$ is the $n \times n$ identity matrix. The $p^n(\lambda)$ can be calculated from the iteration

$$\begin{aligned} p^0(\lambda) &= 1, \\ p^{n+1}(\lambda) &= (\alpha_{n+1} - \lambda)p^n(\lambda) - (\beta_{n-1})^2 p^{n-1}(\lambda). \end{aligned} \tag{7.5}$$

The eigenvalues $T^n$ are the zeros of $p^n(\lambda)$. Thus we wish to find the largest and smallest of these zeros. It can be shown [12] that the number of zeros of $p^n(\lambda)$ lying below some $\lambda$ is given by the number of sign changes in the sequence

$$p^0(\lambda), p^1(\lambda), \ldots p^n(\lambda). \tag{7.6}$$

This relation can then be used to guide a search for the maximum and minimum zeros of $p^n(\lambda)$.

## 8  PRECONDITIONER

With $A$ and $B$ given by $\lambda_{max}$ and $\lambda_{min}$, respectively, the amount of arithmetic required by the algorithm of Sect. 6 is proportional to $\sqrt{\lambda_{max}/\lambda_{min}}$. We now show how the calculation of $tr\log(M^\dagger M)$ can be converted into the calculation of $tr\log(N^\dagger N)$ for an operator $N$ with a smaller value of $\sqrt{\lambda_{max}/\lambda_{min}}$. This change also tends to decrease the number of pseudo-quark fields needed for a reliable evaluation of the trace.

The expression $tr\log(M^\dagger M)$ is gauge invariant since it is $\log det(M^\dagger M)$ and $det(M^\dagger M)$ is gauge invariant. Prior to evaluating $tr\log(M^\dagger M)$ we can therefore transform to a lattice transverse gauge, defined to give a local maximum of the sum over all nearest neighbor $x$ and $y$

$$\sum_{x\,y} tr[u(x,y)]. \tag{8.1}$$

Using, for example, the algorithm described in Ref. [1], the number of arithmetic operations required for gauge fixing is relatively small in comparison to the arithmetic needed to find $\log(M^\dagger M)\phi$ for an ensemble of random pseudo-quark fields $\phi$.

Now define $M_0$ to be the fermion coupling matrix with hopping constant $k_0$ and all $u(x,y)$ equal to 1. Since $M$ has been transformed to a smooth gauge, if the bare gauge coupling constant $g$ is made small and $k_0$ is chosen optimally, we expect $M_0$ to be an approximation to $M$. Thus the preconditioned operator $N$

$$N = (M_0)^{-1} M, \tag{8.2}$$

should be closer to the identity than is $M$. In particular, $\sqrt{\lambda_{max}/\lambda_{min}}$ for $N^\dagger N$ should be smaller than it is for $M^\dagger M$. Using the preconditioned operator we have the relation

$$tr \log(M^\dagger M) = tr \log(N^\dagger N) + tr \log(M_0^\dagger M_0). \tag{8.3}$$

The additional term $tr \log(M_0^\dagger M_0)$ required to find $tr \log(M^\dagger M)$ by Eq. (8.3) does not depend on the gauge configuration and needs to be calculated only once. On the other hand, the operator $M_0$ is diagonal in momentum space. Thus fast fourier transforms provide an efficient way to carry out the multiplication by negative powers of $M_0^\dagger M_0$ needed to determine $\log(N^\dagger N)\phi$ by the algorithm of Sect. 6. A rough guess might be that $\sqrt{\lambda_{max}/\lambda_{min}}$ for $N^\dagger N$ will go to a constant if $g$ is made small, while $\sqrt{\lambda_{max}/\lambda_{min}}$ for $M^\dagger M$ will progressively grow. Thus it seems plausible that for small enough $g$ the additional cost of fourier transforms required to apply the algorithm of Sect. 6 to the preconditioned operator will be more than made up for by the decrease in $\sqrt{\lambda_{max}/\lambda_{min}}$ and corresponding decrease in the number of iterations of Eq. (6.12). At least for the set of parameters at which we run the algorithm in the example describe in the next section, this expectation turns out to be correct.

## 9 EXAMPLE

As a first test, we applied the algorithms of Sects. 6 - 8 to QCD with two flavors of quarks both with k of 0.1600 on a $6^4$ lattice. We calculated a variety of vacuum expectation values using Eq. (3.6) with $n$ of 0, with $n$ of 1, which is the valence approximation, and then found the corresponding error estimates for $n$ of 1 from Eq. (3.7) and corrected vacuum expectations including these error estimates, equivalent to Eq. (5.1) with $n$ of 1 and $m$ of 1. These results were all compared with direct calculations in full QCD.

All calculations in this section were done on an IBM RS/6000 workstation sustaining approximately 10 Mflops. The final production runs with our algorithm required about one week of machine time. The final comparison calculation with full QCD took about another one week. Another month or so of machine time was spent checking and exploring.

As discussed in Section 3, rather than fixing $\beta$, we fixed $\beta'$ given by the sum $\beta + \Delta\beta$. Our first task was then to calculate $\Delta\beta$. From $\beta'$ and $\Delta\beta$ we determined $\beta$ for full QCD. For $\beta'$ we chose the value 5.700. Then $<\ldots>_1$ becomes simply a pure gauge vacuum expectation with pure gauge $\beta'$ of 5.700, and $<\ldots>_0$ is a pure gauge vacuum expectation with the same $\beta$ as used in full QCD, $5.700 - \Delta\beta$.

To evaluate the expectations $<\ldots>_0$ and $<\ldots>_1$, ensembles of pure gauge configurations were generated using the Cabbibo-Marinari-Okawa algorithm. For $<\ldots>_0$ $\beta$ is comparatively small and we were not concerned with obtaining great precision. We found it sufficient to use 100 configurations with 100 sweeps to equilibrate and 100 sweeps between successive pairs. For $<\ldots>_1$, however, we used 160 configurations with 1000 sweeps to equilibrate and 1000 sweeps between successive pairs. For all of the quantities for which $<\ldots>_1$ was measured, we found 1000 sweeps to be more than sufficient to produce equilibrium values and to decorrelate successive values. The full QCD results were found using the hybrid Monte Carlo algorithm. Hamiltonian trajectories were generated using the algorithm of Ref. [11], which is faster than leap-frog by about a factor of 2. Vacuum expectations were taken over an ensemble of 250 accepted trajectories each of length 0.25 time units, with 150 trajectories at $\beta$ of 5.44 followed by 25 trajectories at $\beta$ of 5.439 used to obtain equilibrium.

To tune the algorithm of Sect. 7 for $<\ldots>_1$, we began by evaluating on each gauge configuration the ratio $\sqrt{\lambda_{max}/\lambda_{min}}$ for $M^\dagger M$. For the preconditioned operator $N^\dagger N$, we evaluated this ratio for a range of $k_0$ and found the $k_0$ which minimizes $\sqrt{\lambda_{max}/\lambda_{min}}$. Results are shown in Table 1. The total work required to calculate $\log(M^\dagger M)\phi$ is expected to be about 35% greater than the work required to find $\log(N^\dagger N)\phi$ to the same accuracy. Trial calculations were consistent with this estimate. In the remainder of this section we therefore consider only the preconditioned operator and Eq. (8.3) to find $tr\log(M^\dagger M)$.

Using the preconditioned operator $N$ with the optimal $k_0$, we then calculated the expectation values

$$\begin{aligned} E_1 &= <tr\log(N^\dagger N)>_1 \\ E_2 &= <[tr\log(N^\dagger N)-<tr\log(N^\dagger N)>_1][P-<P>_1]>_1, \end{aligned} \quad (9.1)$$

for a range of different choices of the number of iterations of the Chebyshev algorithm, Eq. (6.12). The results are shown in Table 2. The averages in Table 2 were found using a collection of 16 gauge configurations and 20 random $\phi$ for each configuration. For each configuration, we first calculated $\sqrt{\lambda_{max}/\lambda_{min}}$. The number of iterations of the Chebyshev algorithm, Eq. (6.12), was then chosen to be proportional to $\sqrt{\lambda_{max}/\lambda_{min}}$. The value of $n_{ch}$ shown in Table 2 is the number of iterations which would result for the average over all 160 configurations $<\sqrt{\lambda_{max}/\lambda_{min}}>_1$. As $n_{ch}$ is increased above 50, the change in the two measured expectations shown in Table 2 is significantly less than the statistical errors in the expectations we found with our final, full ensemble of gauge configurations and $\phi$. For the remaining calculations, we chose $n_{ch}$ to be 50.

Using the final ensemble of 160 gauge configurations and a range of values of the number $n_\phi$ of random $\phi$ for each gauge configuration we calculated $E_1$ and $E_2$ of Eq. (9.1). We also evaluated the dispersion $<[P-<P>_1]^2>_1$. We obtained the value $(5.68 \pm 0.61) \times 10^4$. From $E_2$ and $<[P-<P>_1]^2>_1$ we then found $\Delta\beta$. The results are shown in Table 3. As expected, the values we obtained are consistent within errors as $n_\phi$ is varied while the size of the errors themselves tends to fall as $n_\phi$ increases. The optimal choice of $n_\phi$ producing the smallest statistical uncertainty in $\Delta\beta$ for a fixed amount of computation can be shown to be roughly 100.

The error bars on the numbers in these tables are statistical, found by the bootstrap method. From the ensemble of 160 data sets, each consisting of a gauge configuration and an associated collection of $n_\phi$ random $\phi$, we randomly chose 160 new data sets to generate a bootstrap ensemble. On each bootstrap ensemble we then found $E_1$. In this way 100 bootstrap ensembles and 100 values of $E_1$ were found. From these we evaluated the difference between the value of $E_1$ larger than all but 15 results, and the value of $E_1$ smaller than all but 15 results. Half of this difference is shown as the statistical error. The errors for $E_2$ and $\Delta\beta$ were found similarly. In the calculation of the error for $\Delta\beta$, $<[P-<P>_1]^2>_1$ was calculated independently on each bootstrap ensemble and used to determine the corresponding bootstrap value for $\Delta\beta$.

The most reliable value for $\Delta\beta$ is $0.261 \pm 0.014$, obtained with $n_\phi$ of 140. Our algorithm then predicts that expectation values in full QCD with two flavors of quarks, $k$ of 0.1600 and $\beta$ of 5.439 will agree with $<\ldots>_1$ at $\beta'$ given by $\beta + \Delta\beta$ which is $5.700 \pm 0.014$.

Figure 1 shows vacuum expectations of a collection of different Wilson loops. All loops are rectangular with dimensions as shown except for $6_1$ consisting of the 6 link boundary of a pair of orthogonal plaquettes joined on an edge, and $6_2$ consisting of the 6 link boundary of three orthogonal plaquettes joined to form half the surface of a cube. The normalization of each loop is

$$\frac{1}{3}tr[u(x_1,x_2)u(x_2,x_3)\ldots u(x_n,x_1)], \tag{9.2}$$

so that if all link matrices $u(x,y)$ were given by the identity matrix, all loops would become 1. Since the error bars on all points in Figure 1 are smaller than the symbols, in Table 4 we also give numerical values. Each loop expectation shown in the figure is obtained in four different ways. Boxes indicate Eq. (3.6) with $n$ of 0 and $\beta$ of 5.439. Triangles represent the valence approximation, Eq. (3.6) with $n$ of 1 and $\beta + \Delta\beta$ of 5.700. Circles give the valence approximation, Eq. (3.6) with $n$ of 1 and $\beta + \Delta\beta$ of 5.700 but corrected by the error estimate of Eq. (3.7), equivalent to Eq. (5.1) with $n$ of 1 and $m$ of 1. Plusses show full QCD with $\beta$ of 5.439 and two flavors of quarks with $k$ of 0.1600. For all but the $1 \times 1$ loop, clear differences can be seen in Figure 1 between data points with $n$ of 0 and $\beta$ of 5.439 and points for the full theory with two flavors of quarks and $\beta$ of 5.439. For the largest loops these two results differ by as much as a factor of 10. For most loops, nearly all of this shift is correctly reproduced by the valence approximation, $n$ of 1 and $\beta + \Delta\beta$ of 5.700. For the largest loops the valence approximation reproduces most of the shift due to vacuum polarization, but falls noticeably below the results of full QCD. When the leading error estimate of Eq. (3.7) is added to the valence approximation, however, in all cases agreement with full QCD is found within statistical errors. Thus Eq. (3.7) appears to be quite reliable as a error estimate for the valence approximation for the data considered in Figure 1.

A comparison between the numerical efficiency of our algorithm and that of the hybrid Monte Carlo method shows that our method is more efficient for one goal but less efficient for another. For the hybrid Monte Carlo method, we grouped successive configurations into bins and found the averages over each bin. We then evaluated the dispersion in the full ensemble averages by applying the bootstrap method to the binned ensemble. As the size of the bins used in this calculation is made larger, successive bin averages become statistically independent so that the dispersion predicted in this way for the full ensemble average becomes independent of bin size. We determined that about 16 hybrid Monte Carlo trajectories are needed to produce a new statistically independent configuration. On the other hand, since our method uses only a pure gauge updating algorithm, it is relatively inexpensive to guarantee the statistical independence of each successive configuration on which an ensemble of random $\phi$ is constructed. The number of arithmetic operations required to generate one independent configuration by hybrid Monte Carlo turns out to be sufficient to generate about 7 new configurations and $\phi$ ensembles by our method, if the optimal value of $n_\phi$ is chosen.

To obtain a first estimate of a vacuum expectation by either method we might require at least 16 statistically independent configurations be used. Using fewer than perhaps 16 configurations, it is difficult to determine with much confidence whether any independent, equilibrium configurations have been generated. According to this assumption, a first estimate of a vacuum expectation by hybrid Monte Carlo requires about 7 times as much work as by our method. The statistical uncertainty in the hybrid Monte Carlo result found in this way, however, is significantly smaller than that determined by our method. For the data in Figure 1 and Table 4, with approximately equal time spent on full QCD and on the first correction to $< \ldots >_1$ expectations, the statistical errors on the full QCD data are about a factor of 3 smaller than the statistical errors on the sum of $< \ldots >_1$ and its first correction. What the relative performance of the two different algorithms would be for values of lattice spacing, lattice volume and quark mass closer to the continuum limit, we do not yet know.

As a further check of our method, we have calculated the expectation values

$$\begin{aligned} E_3 &= < \{tr[\log(N^\dagger N)] - < tr[\log(N^\dagger N)] >_1\}^2 >_1, \\ E_4 &= < (Q_1)^2 >_1, \\ &= < [\log(N^\dagger N) - L_1]^2 >_1, \\ &= < \{tr\log(N^\dagger N) - < tr\log(N^\dagger N) >_1 - \frac{\Delta\beta}{6}[P - < P >_1]\}^2 >_1, \end{aligned} \qquad (9.3)$$

for $Q_1$ of Eq. (3.5), $L_1$ given by Eq. (3.4) and $\Delta\beta$ of 0.261. A small value of $E_4$ in comparison to $E_3$ suggests that $L_1$ is a good approximation to $tr\log(N^\dagger N)$.

To find these expectations, we evaluated

$$E_5 = \frac{1}{n_\phi} < [\sum_{i=1}^{n_\phi}((\log(N^\dagger N)\phi_i, \log(N^\dagger N)\phi_i))] >_1, \qquad (9.4)$$

$$E_6 = \frac{1}{n_\phi^2} < [\sum_{i=1}^{n_\phi}((\phi_i, \log(N^\dagger N)\phi_i)) - \sum_{i=1}^{n_\phi} < ((\phi_i, \log(N^\dagger N)\phi_i)) >_1]^2 >_1 .$$

As in Section 6, the information that for each $i$ and $x$ the $\phi_i(x)$ are gaussian random variables with covariance given by Eq. (6.2), permits the averages over $\phi$ in the definition of $E_5$ and $E_6$ to be evaluated analytically. It follows that $E_3$ is given by

$$E_3 = E_6 - \frac{E_5}{n_\phi}. \qquad (9.5)$$

We then have

$$E_4 = E_3 - 2\frac{\Delta\beta}{6}E_2 + (\frac{\Delta\beta}{6})^2 < [P - < P >_1]^2 >_1 . \qquad (9.6)$$

Results obtained from 160 gauge field configurations with $n_\phi$ of 140 are given in Table 5 and suggest that $L_1$ is a good approximation to $tr \log(N^\dagger N)$.

A calculation of the effect of quark-antiquark vacuum polarization using a weak coupling perturbation expansion to leading order was reported recently in Ref. [9]. Staggered quarks are considered in Ref. [9] in place of our choice of Wilson quarks. The weak coupling expansion in Ref. [9] is expected to be reliable for sufficiently small gauge coupling and sufficiently large quark mass. For two flavors of quarks with degenerate mass $ma$ in lattice units ranging from 0.05 to 1.00 and a parameter corresponding to $\beta'$ fixed at 5.68, the main effect of quark-antiquark vacuum polarization is found to be simply a coupling constant shift $\Delta\beta$. As $ma$ ranges from 0.05 to 1.00, the shifted $\beta$ for full QCD runs from 5.34 to 5.63.

For Wilson quarks, the mass is

$$ma = \frac{1}{2k} - \frac{1}{2k_c}. \tag{9.7}$$

Here $k_c$ is the critical hopping constant at which the pion mass becomes 0. With our choice of 0.1600 for $k$ and with $k_c$ of 0.1694 [1] at $\beta'$ of 5.700, $ma$ becomes 0.1734. To a first approximation, corresponding versions of QCD with Wilson quarks and with staggered quarks should have equal values of quark mass and $\beta$. Thus the parameters of our trial calculation fall nearly within the range considered in the perturbative calculation of Ref. [9] and our results are qualitatively consistent with theirs. For two flavors of staggered quarks with $ma$ of 0.1734 and $\beta'$ of 5.68, the perturbative calculation predicts a $\Delta\beta$ of 0.226 in comparison to our prediction of $0.261 \pm 0.014$ for two flavors of Wilson quarks.

## 10 CONCLUSION

The crucial question which we have not yet answered is how much time would be required to apply the algorithm we have described to QCD and determine, from Eq. (3.7) with $n$ of 1, the error in the valence approximation to hadron propagators for more realistic choices of quark mass, lattice spacing and lattice volume than we chose in the test in the preceding section. If the algorithm can be run in reasonable time with $n$ of 1, it might be possible to use larger $n$ and obtain smaller errors in hadron propagators. If the error found in this way for some small value of $n$ is itself small for at least one quantity from which the physical value of the lattice spacing can be determined, it will be possible to calculate $\alpha_{\overline{MS}}$ for full QCD.

A perturbation theory estimate, which we will not discuss here, suggests that the optimal number of random $\phi$ which our method requires will grow more slowly than a power of the inverse lattice spacing or the inverse quark mass. Similar estimates suggest similar growth rates for the number of independent gauge configurations needed to evaluate the expectation values entering the determination of the coefficients $a_i$ in the expansion in Eq. (3.1). The remaining question is how large an ensemble of gauge configurations may be required for small values of lattice spacing and quark mass to find the valence approximation error in hadron propagators using Eq. (3.7). If the difference $Q_n$ for some small value of $n$ turns out to be quite small, as occurs for the parameter values in Section 9, or if $Q_n$ is sensitive only to low momentum fluctuations of the gauge field, the calculation of propagator errors may be possible with reasonable ensembles sizes. We do not know at present whether one of these conditions might be realized for values of lattice spacing and quark mass which would permit an extrapolation to the physical limit of hadron masses.

ACKNOWLEDGEMENT

We are grateful to Charles Thorn for several discussions which stimulated much of the work described here.

## A  CONVERGENCE PROOF

We now prove convergence for any bounded $G$ of the sequence of approximate vacuum expectations $<G>_n$ defined by Eq. (3.6). To begin, we will show that convergence holds if the remainders $Q_n$ in Eq. (3.5) have absolute value bounded by a constant independent of $n$ and the gauge field. The required bound on $|Q_n|$ we will then show follows from bounds on the coefficients of $Q_n$ expanded in the basis $\hat{f}_i$ of Section 2, on the vectors $\hat{f}_i$ multiplying these coefficients and on the number of terms occuring in this expansion for any fixed value of the dimension sum $d_i$.

It is convenient to recast Eqs. (2.4) in the form

$$\begin{aligned} <G> &= \frac{Y}{Z} \\ Y &= \zeta^{-1} \int d\mu \, G \, exp(\Delta \mathcal{S} + \mathcal{S}), \\ Z &= \zeta^{-1} \int d\mu \, exp(\Delta \mathcal{S} + \mathcal{S}), \end{aligned} \quad (A.1)$$

and Eqs. (3.6) in the form

$$\begin{aligned} <G>_n &= \frac{Y_n}{Z_n} \\ Y_n &= \zeta^{-1} \int d\mu \, G \, exp(\Delta \mathcal{S} - Q_n + \mathcal{S}), \\ Z_n &= \zeta^{-1} \int d\mu \, exp(\Delta \mathcal{S} - Q_n + \mathcal{S}), \end{aligned} \quad (A.2)$$

where $\mathcal{S}$ is the bounded effective action entering the inner product of Eq. (2.2), $\zeta$ is defined in Eqs. (2.2), $Q_n$ is defined by Eq. (3.5) and $\Delta \mathcal{S}$ is given by

$$\Delta \mathcal{S} = \log det(M^\dagger M + R) + \frac{\beta}{6} P - \mathcal{S}. \quad (A.3)$$

The difference $Y - Y_n$ can be expressed in the form

$$Y - Y_n = \zeta^{-1} \int d\mu \, G \, Q_n \, \frac{1 - exp(-Q_n)}{Q_n} \, exp(\Delta \mathcal{S} + \mathcal{S}). \quad (A.4)$$

The discussion of Sections 2 and 3 combined with the boundedness of $\mathcal{S}$ implies $\Delta \mathcal{S}$ is bounded in absolute value. We will show below that $Q_n$ is bounded in absolute value by a constant independent of $n$ and the gauge field. It follows that $[1 - exp(-Q_n)]/Q_n$ is bounded in absolute value by a constant independent of $n$ and gauge field. Thus from Eq. (A.4) it follows that there is a constant $c$ independent of $n$ such that

$$|Y - Y_n| \leq c \, \zeta^{-1} \int d\mu \, |G| \, |Q_n| \, exp(\mathcal{S}). \quad (A.5)$$

By the Cauchy-Schwarz inequality we then have

$$|Y - Y_n| \leq c \, \|G\| \, \|Q_n\|, \tag{A.6}$$

where $\|\ldots\|$ is the norm defined by Eq. (2.1). The discussion of Section 3 implies $\|Q_n\|$ goes to 0 as $n \to \infty$. Thus $|Y - Y_n|$ goes to 0. The preceding argument with $G$ chosen to be 1 implies that $|Z - Z_n|$ goes to 0. Thus by Eqs. (A.1) and (A.2), $<G>$ approaches $<G>_n$.

To show that $Q_n$ is bounded in absolute value by a constant independent of $n$ and the gauge field, it is easier to work with a related remainder $Q'_n$ found by expanding $\log det(M^\dagger M + R)$ using the basis $\hat{f}_i$ of $\mathcal{F}$. In $\mathcal{F}$ we have

$$\log det(M^\dagger M + R) = \sum_i b_i \hat{f}_i, \tag{A.7}$$

$$b_i = (\hat{f}_i, \log det(M^\dagger M + R)). \tag{A.8}$$

Define the remainder for truncations of this expansion $Q'_n$ to be

$$Q'_n = \sum_{i > n} b_i \hat{f}_i. \tag{A.9}$$

Each $\hat{h}_i$ in the basis for $\mathcal{H}$ is a linear combination of $\hat{f}_j$ in the basis for $\mathcal{F}$ with a single fixed value of total dimension $d_j$. In addition, distinct $\hat{h}_i$ and $\hat{h}_{i'}$ are linear combinations of disjoint sets of $\hat{f}_j$. It follows that there is a way of choosing the details of the ordering of basis vectors $\hat{f}_j$ which we specified in Section 2 which makes the remainder sequence $Q_n$ a subsequence of the remainder sequence of $Q'_n$. It is therefore sufficient for us to establish the existence of a bound on $|Q'_n|$ independent of $n$ and the gauge field.

## B  EXPANSION COEFFICIENT BOUND

As discussed in Section 3, since $M$ is a finite matrix with matrix elements bounded uniformly over all gauge fields, the spectrum of $M^\dagger M + R$ is bounded from above by some constant $A$. The spectrum of $M^\dagger M + R$ is also bounded from below by $R$. We can therefore express $\log det(M^\dagger M + R)$ in the form

$$\log det(M^\dagger M + R) = \log det(\frac{M^\dagger M}{A} + \frac{R}{A}) + d_M \log(A), \tag{B.1}$$

$$= tr \log[1 - (1 - \frac{M^\dagger M}{A} - \frac{R}{A})] + d_M \log(A). \tag{B.2}$$

where $d_M$ is the dimension of the matrix $M$. The spectrum of the matrix $1 - \frac{M^\dagger M}{A} - \frac{R}{A}$ is nonnegative and bounded from above by $1 - \frac{R}{A}$ independent of the gauge fields. Thus the logarithm in Eq. (B.2) can be expanded as a power series,

$$\log det(M^\dagger M + R) = -\sum_j \frac{1}{j} tr[(1 - \frac{M^\dagger M}{A} - \frac{R}{A})^j] + d_M \log(A), \tag{B.3}$$

which converges as a result of the bound, independent of gauge field,

$$tr[(1 - \frac{M^\dagger M}{A} - \frac{R}{A})^j] \le d_M (1 - \frac{R}{A})^j. \tag{B.4}$$

The convergence bound Eq. (B.4) permits Eq. (B.3) to be substituted into Eq. (A.8) to obtain an expansion for the coefficients $b_i$,

$$b_i = \sum_j b_{ij},$$

$$b_{ij} = -\frac{1}{j}(\hat{f}_i, tr[(1 - \frac{M^\dagger M}{A} - \frac{R}{A})^j]). \tag{B.5}$$

From this expansion we obtain a bound on the $b_i$.

The matrix elements of $M^\dagger M$ are either independent of the gauge field or are linear functions of matrix elements of the product $u(x,y)u(y,z)$ for a pair of adjoining nearest neighbor links $(x,y)$ and $(y,z)$. Thus $(1 - \frac{M^\dagger M}{A} - \frac{R}{A})^j$ includes products of at most $j$ matrix elements of $u(x,y)u(y,z)$. For the trace $tr[(1 - \frac{M^\dagger M}{A} - \frac{R}{A})^j]$ written as a linear combination of the irreduceable representations $f_i$ discussed in Section 2, we will show that the dimension sums $d_i$ which occur are bounded by

$$d_i < \ell + \frac{1}{4}(j+3)^3, \tag{B.6}$$

where $\ell$ is the number links in the lattice.

Standard results on products of $SU(3)$ representations imply that for a single link the largest dimension $d$ of an irreducable representation which can occur in the product of $p_1$ copies of matrix elements of $u(x,y)$ and $p_2$ copies of matrix elements of the conjugate matrix $u(y,x)$ corresponds to a Young tableau with $p_1$ columns containing one box and $p_2$ columns containing two boxes. The dimension $d$ is then given by

$$d = \frac{1}{2}(p_1 + 1)(p_2 + 1)(p_1 + p_2 + 2). \tag{B.7}$$

Now consider the dimension sums $d_i$ which occur if a product of $j$ matrix elements of $u(x,y)u(y,z)$ for any collection of $j$ pairs of adjoining links $(x,y)$ and $(y,z)$ is expressed as a sum of the products of irreduceable representations $f_i$. Since $d$ in Eq. (B.7) rises faster than linearly in $p_1$ and $p_2$, the dimension sum $d_i$ will be greatest if all $j$ matrices $u(x,y)u(y,z)$ are taken on a single pair of adjoining links. By Eq. (B.7) this will be maximized if $p_1$ matrix elements of $u(x,y)u(y,z)$ are chosen and $p_2$ elements of the conjugate matrix $u(z,y)u(y,x)$ are chosen with $p_1$ given by the largest integer less than or equal to $j/2$ and $p_2$ is given by $j - p_1$. The contribution to $d_i$ coming from these two links is $2d$ for $d$ given by Eq. (B.7). This sum is bounded by the second term in Eq. (B.6). The remaining lattice links all carry the trivial 1-dimensional representation giving a sum bounded by the first term in Eq. (B.6).

The $f_i$, however, have been ordered with increasing values of $d_i$, and the $\hat{f}_i$ come from a Gram-Schmidt process applied to the $f_i$. Thus $\hat{f}_{i'}$ is orthogonal to any $f_i$ with $d_i$ less than $d_{i'}$. Therefore $b_{ij}$ in Eq. (B.5) vanishes unless $j$ obeys Eq. (B.6). The Cauchy-Schwarz inequality applied to Eq. (B.5), combined with Eq. (B.4) and (B.6) then gives the bound

$$|b_i| < c(1 - \frac{R}{A})^{(c' d_i^{\frac{1}{3}})}, \tag{B.8}$$

for a pair of constants $c$, $c'$ independent of $i$. For the infinite set of $i$ large enough that $d_i$ is much larger than $\ell$, the $\frac{1}{3}$ power in Eq. (B.8) follows from the power 3 in Eq. (B.6). By choosing $c$ sufficiently large, Eq. (B.8) can then be made to hold also for the remaining finite set of smaller $i$ and $d_i$.

## C  BASIS VECTOR BOUND

We next derive a bound on the $\hat{f}_i$ independent of the gauge field. This bound is obtained by examining the Gram-Schmidt process by which the sequence of $\hat{f}_i$ is constructed from the sequence of $f_i$.

The vector $\hat{f}_i$ can be found by normalizing the vector $p_i$

$$\hat{f}_i = \frac{p_i}{\|p_i\|}, \tag{C.1}$$

$$p_i = f_i + \sum_{j<i} \alpha_{ij} f_j, \tag{C.2}$$

with coefficients $\alpha_{ij}$, defined only for $j$ less than $i$, chosen to minimize the norm

$$\|p_i\|^2 = (f_i + \sum_{j<i} \alpha_{ij} f_j, f_i + \sum_{j<i} \alpha_{ij} f_j). \tag{C.3}$$

By the triangle inequality and Eq. (C.2), we have for any gauge field and any $i$

$$|p_i| \leq |f_i| + \sum_{j<i} |\alpha_{ij}||f_j|. \tag{C.4}$$

The $f_j$ are normalized with respect to pure Haar measure or, equivalently, using the inner produce of Eq. (2.2) with $\mathcal{S}$ replace by 0. For any irreducable representation of $SU(3)$ by unitary matrices $v_{ab}$, of dimension $d \times d$, we have for an integral over Haar measure $\nu$ on a single copy of $SU(3)$

$$\int v_{ab}^* v_{ab} \, d\nu = \frac{1}{d}. \tag{C.5}$$

Thus the matrix $\sqrt{d} v_{ab}$ is normalized to 1 and has matrix elements bounded by $\sqrt{d}$. Applying this argument to the copy of $SU(3)$ on each lattice link and then taking a maximum of the resulting product of dimensions we obtain the bound, independent of gauge field,

$$|f_j| \leq (\frac{d_j}{\ell})^{\frac{\ell}{2}}. \tag{C.6}$$

Eqs. (C.2) and (C.6), combined with the Cauchy-Schwarz inequality and the information that the sequence of $d_j$ is nondecreasing then imply for any gauge field

$$|p_i| \leq (1 + \sum_{j<i} |\alpha_{ij}|^2)^{\frac{1}{2}} (\frac{d_i}{\ell})^{\frac{\ell}{2}}. \tag{C.7}$$

Now return to the squared norm $(p_i, p_i)$. Since $\mathcal{S}$ in Eq. (2.2) defining the norm is bounded from above and below, and since the $f_j$ are orthonormal with respect to Haar measure, we have for a constant $c$ independent of $i$

$$\begin{aligned}(p_i, p_i) &= \zeta^{-1} \int d\mu |f_i + \sum_{j<i} \alpha_{ij} f_j|^2 exp(\mathcal{S}), &\text{(C.8)}\\ &\geq c \int d\mu |f_i + \sum_{j<i} \alpha_{ij} f_j|^2 &\text{(C.9)}\\ &= c(1 + \sum_{j<i} |\alpha_{ij}|^2). &\text{(C.10)}\end{aligned}$$

Eqs. (C.1), (C.7) and (C.8) then yield for some constant $c'$ independent of $i$ and the gauge field

$$|\hat{f}_i| \leq c' (\frac{d_i}{\ell})^{\frac{\ell}{2}}. \tag{C.11}$$

## D   REMAINDER BOUND

To complete the bound on the remainders $Q'_n$ we need a bound on the number $n(d)$ of $\hat{f}_i$ entering Eq. (A.9) for any fixed value $d$ of the dimension sum $d_i$. The number $n(d)$ of such $\hat{f}_i$ is certainly bounded by

$$n(d) \leq [d^2 m(d)]^\ell, \tag{D.1}$$

where $m(d)$ is the number of distinct irreducable representation of dimension $d$ or less for a single copy of $SU(3)$ and $d^2$ is clearly an upper bound on the number of matrix elements in each such matrix. Each irreduceable representation of dimension $d$ or less corresponds to a Young tableau specified by $p_1$ and $p_2$ as before but with Eq. (B.7) now replaced by an inequality

$$d \geq \frac{1}{2}(p_1 + 1)(p_2 + 1)(p_1 + p_2 + 2). \tag{D.2}$$

Any $p_1$ or $p_2$ fulfilling Eq. (D.2) is less than $\sqrt{2d}$. Thus the total number of such pairs is less than $2d$. Eq. (D.1) becomes

$$n(d) < (2d^3)^\ell. \tag{D.3}$$

Combining Eq. (A.9) for the remainder $Q_n$ with the bounds of Eqs. (B.8), (C.11) and (D.3) we obtain a bound of the form

$$|Q'_n| \leq \sum_{d \geq d_n} c'' d^{\frac{7\ell}{2}} (1 - \frac{R}{A})^{(c'd^{\frac{1}{3}})}. \tag{D.4}$$

The sum in Eq. (D.4) is over all integers $d$ greater than or equal to the dimension sum $d_n$ for $\hat{f}_n$. The parameters $R$, $A$, $c$ and $c''$ are independent of $d$ and the gauge field. Since $d_n$ is a positive nondecreasing function of $n$, the sum in Eq. (D.4) is less than the sum with $d_n$ replaced by 0, which in turn is convergent and gives the required bound on $Q'_n$ independent of both $n$ and the gauge field.

## E  CHEBYSHEV EXPANSION

We now derive the coefficients $n_i$ needed for the Chebyshev expansion of $\log y$ Eq. (6.7). From standard results on Chebyshev polynomials [13], it follows that

$$\frac{1}{y}[1 + \rho T^*_{n+1}(\frac{1-y}{1-\epsilon})] = \sum_{k=0}^{n} c_k T^*_k(\frac{1-y}{1-\epsilon}), \tag{E.1}$$

where for $1 \leq k \leq n$

$$c_k = 2\frac{(1 + \cosh\chi)\sinh[(n+1-k)\chi]}{\sinh\chi \cosh[(n+1)\chi]}, \tag{E.2}$$

and in addition

$$c_0 = \frac{(1 + \cosh\chi)\sinh[(n+1)\chi]}{\sinh\chi \cosh[(n+1)\chi]},$$

$$\rho = \frac{-1}{\cosh[(n+1)\chi]}, \tag{E.3}$$

with $\chi$ defined by

$$\cosh \chi = \frac{1+\epsilon}{1-\epsilon}. \tag{E.4}$$

Eq. (E.1) can be integrated from $y$ to 1 using, for $k$ greater than 1, the relation

$$\int T_k^*(x)dx = \frac{T_{k+1}^*(x)}{4(k+1)} - \frac{T_{k-1}^*(x)}{4(k-1)}, \tag{E.5}$$

along with

$$\int T_1^*(x)dx = \frac{T_2^*(x)}{8} - \frac{T_0^*(x)}{8},$$
$$\int T_0^*(x)dx = \frac{T_1^*(x)}{2} + \frac{T_0^*(x)}{2}. \tag{E.6}$$

We obtain

$$\log y = \sum_{k=0}^{n+1} b_k T_k^*\left(\frac{1-y}{1-\epsilon}\right) + \delta \log y \tag{E.7}$$

where for $1 \leq k \leq n$

$$b_k = -\frac{(1-\epsilon)(1+\cosh\chi)\cosh[(n+1-k)\chi]}{k\cosh[(n+1)\chi]}, \tag{E.8}$$

and in addition

$$b_{n+1} = -\frac{(1-\epsilon)(1+\cosh\chi)}{2(n+1)\cosh[(n+1)\chi]},$$
$$b_0 = -\sum_{k=1}^{n+1}(-1)^k b_k. \tag{E.9}$$

The bound Eq. (6.8) on $\delta$ follows from the integral of Eq. (E.1) combined with Eq. (E.2) for $\rho$ and the bound

$$|T_k^*(x)| \leq 1, \tag{E.10}$$

for $0 \leq x \leq 1$.

| $\mathcal{O}$ | $k_0$ | $<\sqrt{\lambda_{max}/\lambda_{min}}>$ | work/sweep | work |
|---|---|---|---|---|
| $M^\dagger M$ |  | 58.1 | 1.0 | 58.1 |
| $N^\dagger N$ | 0.091 | 22.1 | 2.0 | 44.2 |

Table 1: Comparison of work required to evaluate $\log(\mathcal{O})\phi$ for different choices of $\mathcal{O}$

| $n_{ch}$ | $E_1$ | $E_2$ |
|---|---|---|
| 10 | 390.22 | 6870 |
| 20 | 412.25 | 2733 |
| 50 | 412.74 | 2843 |
| 100 | 412.79 | 2848 |
| 200 | 412.79 | 2848 |

Table 2: Expectation values found from 16 gauge configurations each with 20 random $\phi$, for various choices of the number of iterations of the Chebyshev algorithm used in the calculation of $\log(N^\dagger N)\phi$.

| $n_\phi$ | $E_1$ | $E_2$ | $\Delta\beta$ |
|---:|---|---|---|
| 10 | $416.8 \pm 2.1$ | $2340 \pm 508$ | $0.247 \pm 0.049$ |
| 20 | $416.0 \pm 1.7$ | $2343 \pm 387$ | $0.248 \pm 0.029$ |
| 30 | $415.8 \pm 1.5$ | $2483 \pm 367$ | $0.262 \pm 0.028$ |
| 40 | $415.4 \pm 1.4$ | $2531 \pm 343$ | $0.267 \pm 0.026$ |
| 50 | $414.7 \pm 1.3$ | $2517 \pm 344$ | $0.266 \pm 0.024$ |
| 60 | $414.8 \pm 1.3$ | $2499 \pm 318$ | $0.264 \pm 0.023$ |
| 70 | $414.9 \pm 1.2$ | $2484 \pm 309$ | $0.262 \pm 0.020$ |
| 80 | $414.7 \pm 1.2$ | $2470 \pm 294$ | $0.261 \pm 0.018$ |
| 90 | $414.7 \pm 1.1$ | $2420 \pm 290$ | $0.256 \pm 0.017$ |
| 100 | $414.8 \pm 1.1$ | $2483 \pm 307$ | $0.262 \pm 0.017$ |
| 110 | $414.8 \pm 1.1$ | $2523 \pm 309$ | $0.267 \pm 0.015$ |
| 120 | $414.5 \pm 1.0$ | $2501 \pm 296$ | $0.264 \pm 0.014$ |
| 130 | $414.2 \pm 1.0$ | $2489 \pm 290$ | $0.263 \pm 0.014$ |
| 140 | $414.5 \pm 1.0$ | $2472 \pm 288$ | $0.261 \pm 0.014$ |

Table 3: Expectation values and $\Delta\beta$ found from 160 gauge configurations for various choices of the number of random $\phi$ used in the evaluation of traces.

| loop | $n=0$ $\beta = 5.439$ | valence $n=1$ $\beta' = 5.700$ | valence + error $n=1, m=1$ $\beta' = 5.700$ | full QCD $\beta = 5.439$ $\kappa = .16$ |
|---|---|---|---|---|
| $1 \times 1$ | 0.4827 (05) | 0.5501 (04) | 0.5501 (29) | 0.5527 (11) |
| $2 \times 1$ | 0.2438 (06) | 0.3261 (06) | 0.3325 (44) | 0.3341 (16) |
| $6_1$ | 0.2747 (06) | 0.3618 (05) | 0.3655 (38) | 0.3678 (16) |
| $6_2$ | 0.2212 (06) | 0.3143 (06) | 0.3208 (42) | 0.3230 (17) |
| $3 \times 1$ | 0.1243 (05) | 0.1966 (06) | 0.2036 (46) | 0.2064 (15) |
| $2 \times 2$ | 0.0678 (05) | 0.1341 (06) | 0.1479 (46) | 0.1471 (16) |
| $4 \times 1$ | 0.0635 (04) | 0.1188 (05) | 0.1249 (43) | 0.1283 (14) |
| $5 \times 1$ | 0.0326 (03) | 0.0720 (04) | 0.0774 (35) | 0.0799 (10) |
| $3 \times 2$ | 0.0195 (03) | 0.0592 (04) | 0.0713 (32) | 0.0706 (13) |
| $4 \times 2$ | 0.0057 (02) | 0.0264 (03) | 0.0346 (24) | 0.0350 (09) |
| $3 \times 3$ | 0.0038 (03) | 0.0206 (03) | 0.0307 (26) | 0.0288 (09) |

Table 4: Vacuum expectation of various Wilson loops found by four different methods.

| | |
|---|---|
| $< [tr[\log(N^\dagger N)] - < tr[\log(N^\dagger N)] >_1]^2 >_1$ | $124 \pm 16$ |
| $< [tr \log(N^\dagger N) - L_1]^2 >_1$ | $16.0 \pm 6.3$ |

Table 5: Expectation values measuring how good an approximation $L_1$ is to $tr \log(N^\dagger N)$. The calculation uses 160 gauge configurations each with 140 random $\phi$.

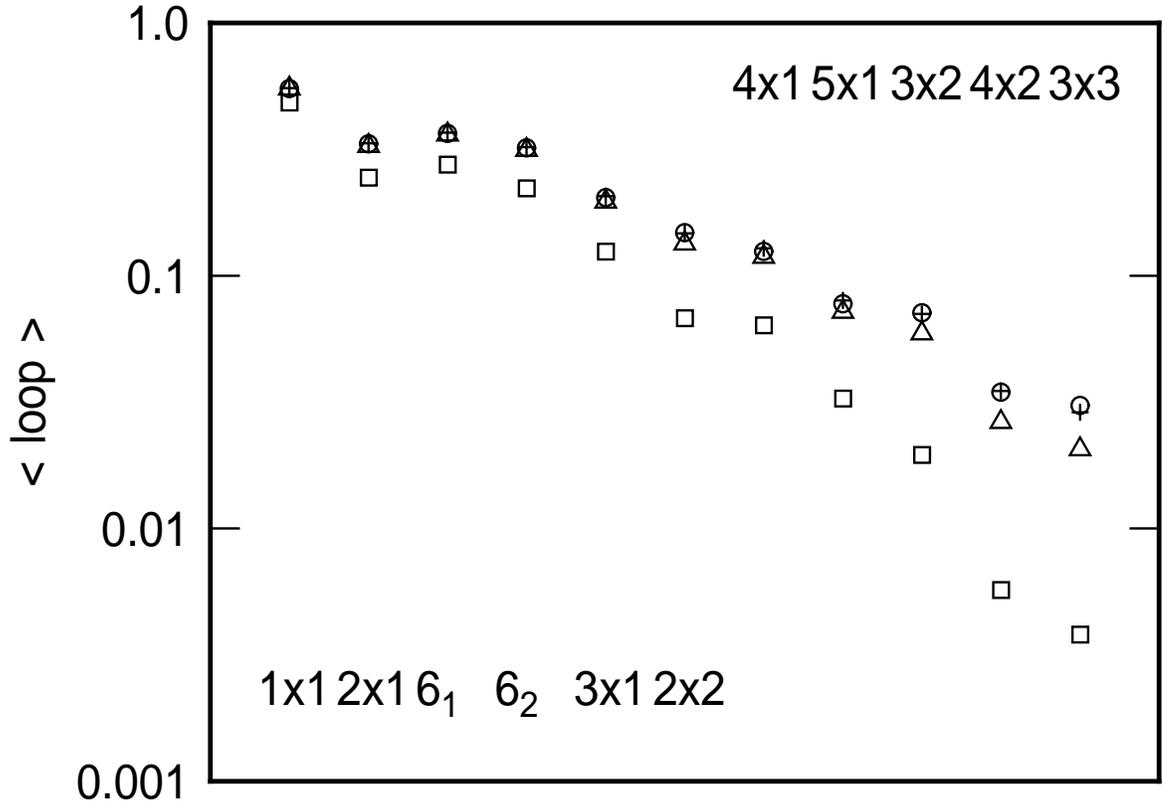

Figure 1: Vacuum expectations of various Wilson loops found by four different methods. Boxes represent $n = 0$ with $\beta = 5.439$, triangles are the valence approximation $n = 1$ with $\beta' = 5.700$, circles are the valence approximation plus first error estimate, equivalent to $n = 1$, $m = 1$ with $\beta' = 5.700$, plusses are full QCD with $\beta = 5.439$ and two quark flavors with $k = 0.1600$.

37